\documentclass[prd,twocolumn,amsmath,amssymb,floatfix, superscriptaddress,nofootinbib]{revtex4}

\usepackage{bm}
\usepackage{amsmath}
\usepackage{epsf}
\usepackage{color}
\usepackage{natbib}
\usepackage{graphicx}

\makeatletter
\renewcommand\footnoterule{%
  \vspace{-5pt}
  \kern-3\p@\hrule\@width.4\columnwidth%
  \kern10\p@}
\makeatother

\def\be{\begin{equation}}
\def\ee{\end{equation}}
\def\ba{\begin{eqnarray}}
\def\ea{\end{eqnarray}}

\def\mnras{MNRAS}

\begin{document}

\title{The Shape of the Primordial Power Spectrum: A Last Stand Before Planck}
\author{Hiranya V. Peiris}\email{h.peiris@ucl.ac.uk}
\affiliation{Institute of Astronomy and Kavli Institute for Cosmology, University of Cambridge, Cambridge CB3 0HA, U.K.}
\affiliation{Department of Physics and Astronomy, University College London, London WC1E 6BT, U.K.}

\author{Licia Verde}\email{liciaverde@icc.ub.edu}
\affiliation{ICREA \& Instituto de Ciencias del Cosmos, Universitat de Barcelona, Marti i Franques 1, 08028, Barcelona, Spain}

\date{\today}

\begin{abstract}
We present a minimally-parametric reconstruction of the primordial power spectrum using the most recent cosmic microwave background and large scale structure data sets. Our goal is to constrain the shape of the power spectrum while simultaneously avoiding strong theoretical priors and over-fitting of the data. We find no evidence for any departure from a power law spectral index. We also find that an exact scale-invariant power spectrum is disfavored by the data, but this conclusion is 
 weaker than the corresponding result assuming a theoretically-motivated power law spectral index prior. The reconstruction shows that better data are crucial to justify the adoption of such a strong theoretical prior observationally. These results can be used to determine the robustness of our present knowledge when compared with forthcoming precision data from Planck.
\end{abstract}
\maketitle

\section{Introduction}\label{sec:intro}

The deviation from scale invariance of the primordial scalar power spectrum is a critical prediction of inflation, and unlike other potential signatures such tensor modes or non-Gaussianity,
it is the only signature that is {\it generic to all inflationary models}. It is therefore a vital test of the inflationary paradigm, and we address it with a minimally parametric approach.

Briefly, the idea is as follows. Choose a functional form which allows a great deal of freedom in the form of the deviation from scale invariance ({\it e.g.} smoothing splines). Naively fitting this to the data will lead one to fit the fluctuations due to cosmic variance and experimental noise, with arbitrary improvement in the chi-square. Instead, one performs cross-validation: throw out some of the data (the validation set), fit the rest (the training set), and see how well it predicts the validation set. A very good fit to the training set, which poorly predicts the validation set, indicates over-fitting of noisy data. The final ingredient in the algorithm is a roughness penalty, a parameter that penalizes a high degree of structure in the functional form. By performing cross-validation as a function of this penalty, one can judge when the amount of freedom in the smoothing spline is what the data require without fitting the noise. 
A minimally parametric power spectrum reconstruction  combined with a  roughness penalty set by  cross-validation  thus provides a method of determining smooth departures from scale invariance which avoids two pitfalls. Firstly, a strong theory prior on the form of the power spectrum ({\it e.g.} the commonly used power law prescription) can lead to artificially tight constraints on -- even a spurious detection of -- a deviation from scale invariance, which is mostly due to the strength of the prior than that of the data. Secondly, simple binning techniques \cite{Bridleetal03,SpergelWMAP06,Hannestad04,BridgesLasenbyHobson06,BridgesLasenbyHobson07} or direct inversion \cite{ShafielooSouradeep04, ShafielooSouradeep07, Kogoetal04,Tocchini-Valentinietal06, Tocchini-Valentinietal05, nagata08,nagata09,Ichiki09, Nicholson09a, Nicholson09b} of the data to obtain the primordial power spectrum can lead one to fit noisy data with a large improvement in chi-square\footnote{Given a calibration uncertainty in the covariance matrix leading to an uncertainty of 3\% in the absolute calibration of $\chi^2$/d.o.f. (easily plausible with current data), it would not be surprising to see an improvement of $\Delta \chi^2 \sim -30$ relative to a smooth power spectrum by ``fitting the noise'' with a power spectrum containing a high degree of structure.}. A minimally parametric approach combined with  cross-validation avoids these issues, providing a way to actually determine the strength of the shape prior justified by the quality of the data. Cross-validation would also be helpful for alternative minimally-parametric methods \cite{MukherjeeWang05,Leach05}  {\it e.g.} in choosing the number of basis functions.

In this work, we use  the best available data over a wide range of scales corresponding to the longest ``lever arm'' of wavenumbers currently extant, to reconstruct the shape of the primordial power spectrum in a minimally-parametric way.
ESA's Planck satellite, which has already begun taking data, is expected to provide superior constraints \cite{PlanckBlue} on the shape of the primordial scalar power spectrum by 2012. Our goal here is to establish a benchmark of what was known about the shape of the power spectrum before the Planck analysis.

\begin{figure*}[!htp]
\includegraphics[scale=0.5]{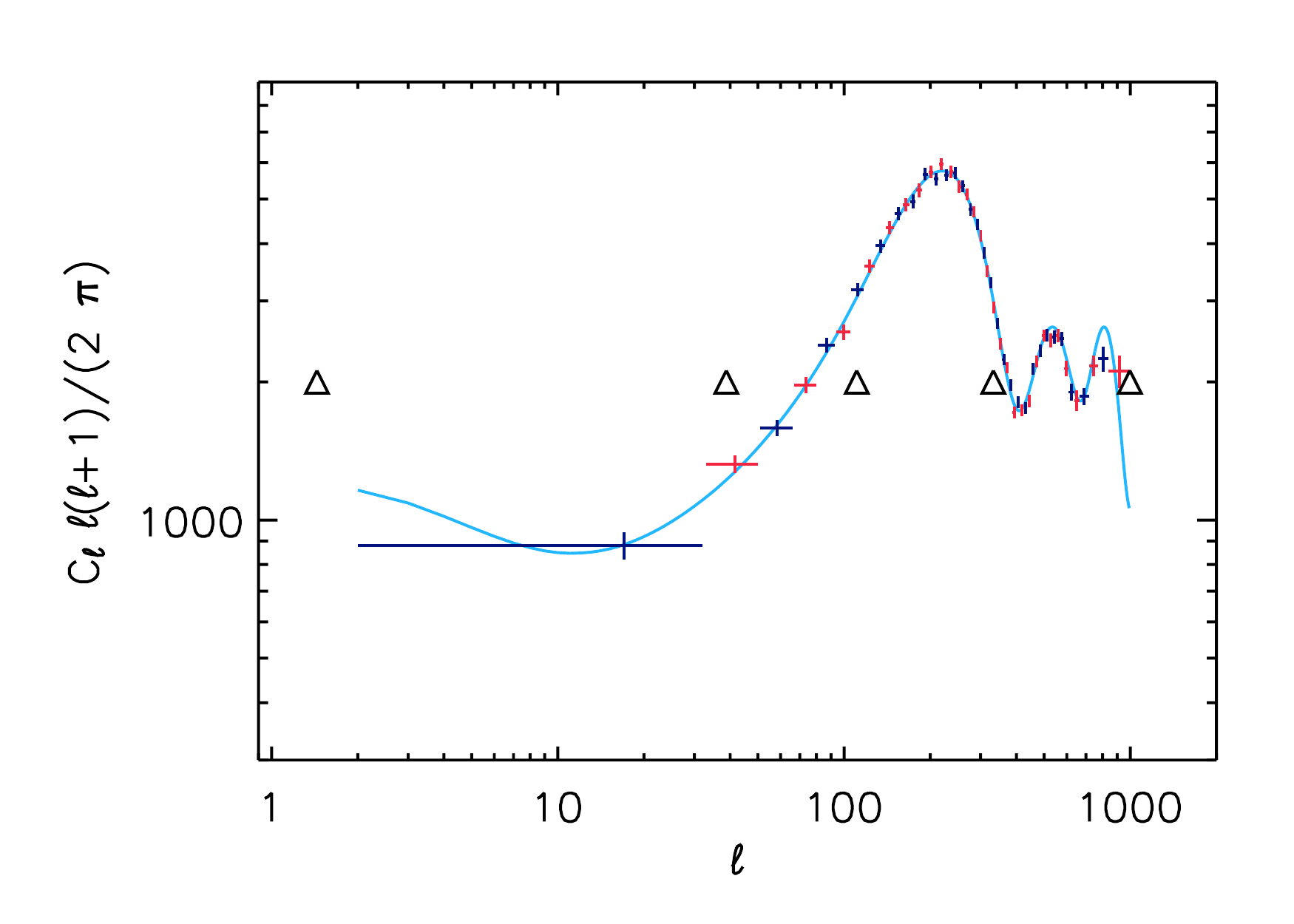}
\includegraphics[scale=0.5]{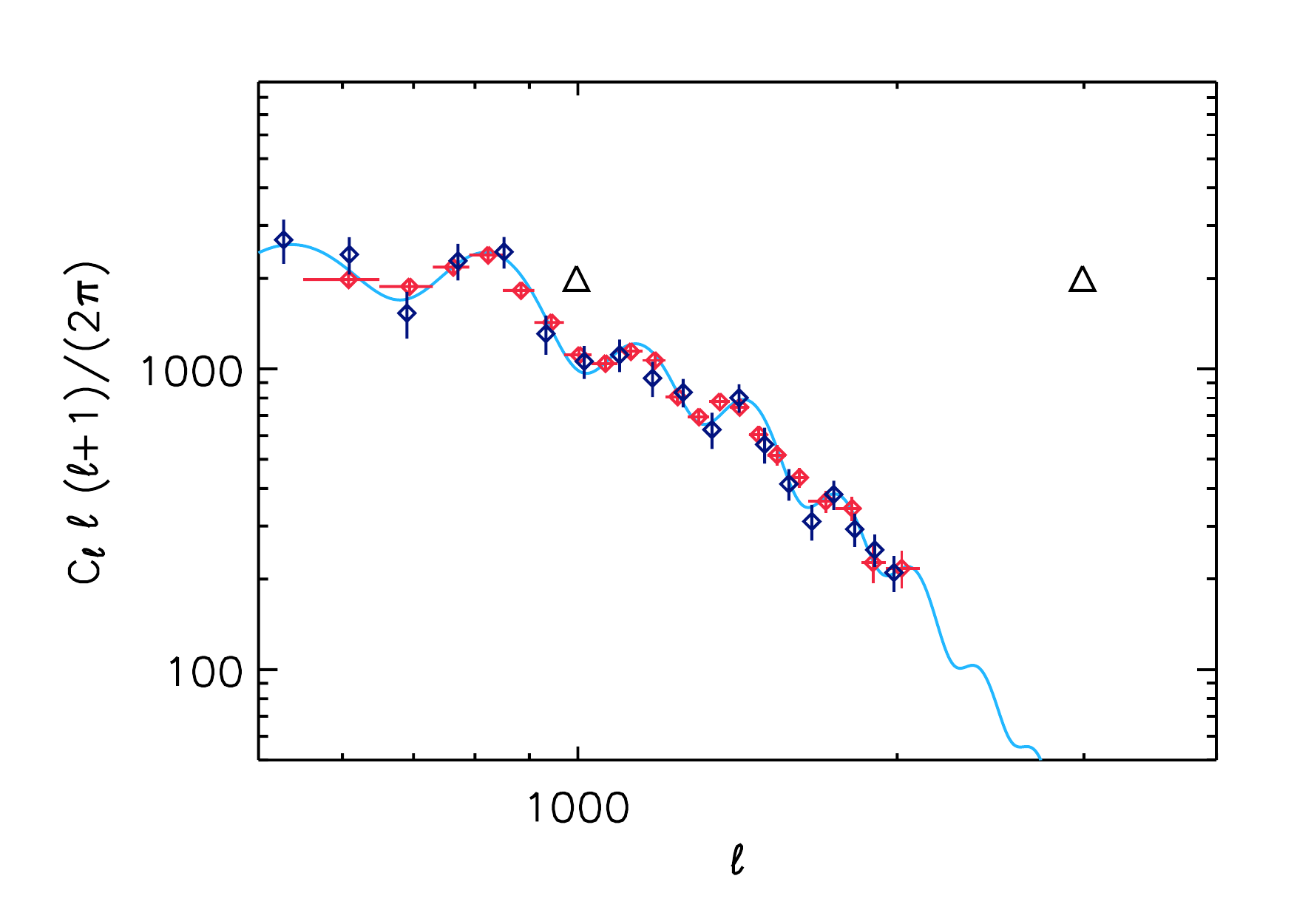}
\caption{(Left) WMAP 5 year data \cite{NoltaWMAP5} and (right) external CMB data from ACBAR and QUaD \cite{ACBAR09,Quad09}, showing knot placement (triangles, arbitrary normalization) and the cross-validation set-up. CV$_A$ is red and CV$_B$ is blue. We show only the temperature data here (in $\mu$K$^2$) as the constraints on the power spectrum shape come mostly from the temperature data; in practice for each data set  we also use the polarization data, which is crucial in lifting degeneracies with the cosmological parameters. The light blue line is a concordance LCDM model.} 
\label{fig:cvcmb}
\end{figure*}

\section{Methodology}\label{sec:methods}
We perform  a minimally-parametric reconstruction of the primordial power spectrum based on the method of  Ref.~\cite{Sealfon05}. Since the simplest inflationary models, which are consistent with the data, predict  the primordial power spectrum to be a smooth function, we search for smooth deviations\footnote{A Bayesian reconstruction technique has been proposed in Ref. \cite{Bridges09} which also avoids overfitting of the data and is perhaps more suited for discovering local violations of scale invariance.} from scale invariance with a cubic smoothing spline technique (for details, see Refs.~\cite{GreenSilverman,Sealfon05,VerdePeiris08} which we only briefly summarize here). In this approach, one aims to recover a function $f(x)$ from measurements $\hat{f}$ at $n$ discrete points $x_i$. 

Consider a description of $f$ by a piecewise cubic spline $F(x)$. It is uniquely defined by the values of $F$ at $N$ ``knots'' once we ask for continuity of $F(x)$ and its first and second derivatives at the knots, and two boundary conditions: we require the second derivative to vanish at the exterior knots. In our application, $F(x)$ is the primordial power spectrum $P(k)$, and the data are: the angular power spectrum of the 5 year Wilkinson Microwave Anisotropy Probe (WMAP5) cosmic microwave background (CMB) temperature and polarization \cite{NoltaWMAP5}; alone or in combination with higher resolution, ground-based CMB experiments (QUaD \cite{Quad09} and ACBAR \cite{ACBAR09}); or with large scale structure data: the Sloan Digital Sky Survey (SDSS) Data Release 7 (DR7) Luminous Red Galaxy (LRG) power spectrum \cite{Reidetal09}; and the Lyman-alpha forest (Ly$\alpha$) power spectrum constraints from Ref. \cite{McDonaldLya06}.  This work thus represents a significant advance over previous work \cite{VerdePeiris08}, with a new WMAP release (two further years of data and a significant advance in the understanding of systematic errors) plus substantial improvements in both ground-based CMB data and large-scale structure data. 

\begin{figure}[!t]
\includegraphics[scale=0.48]{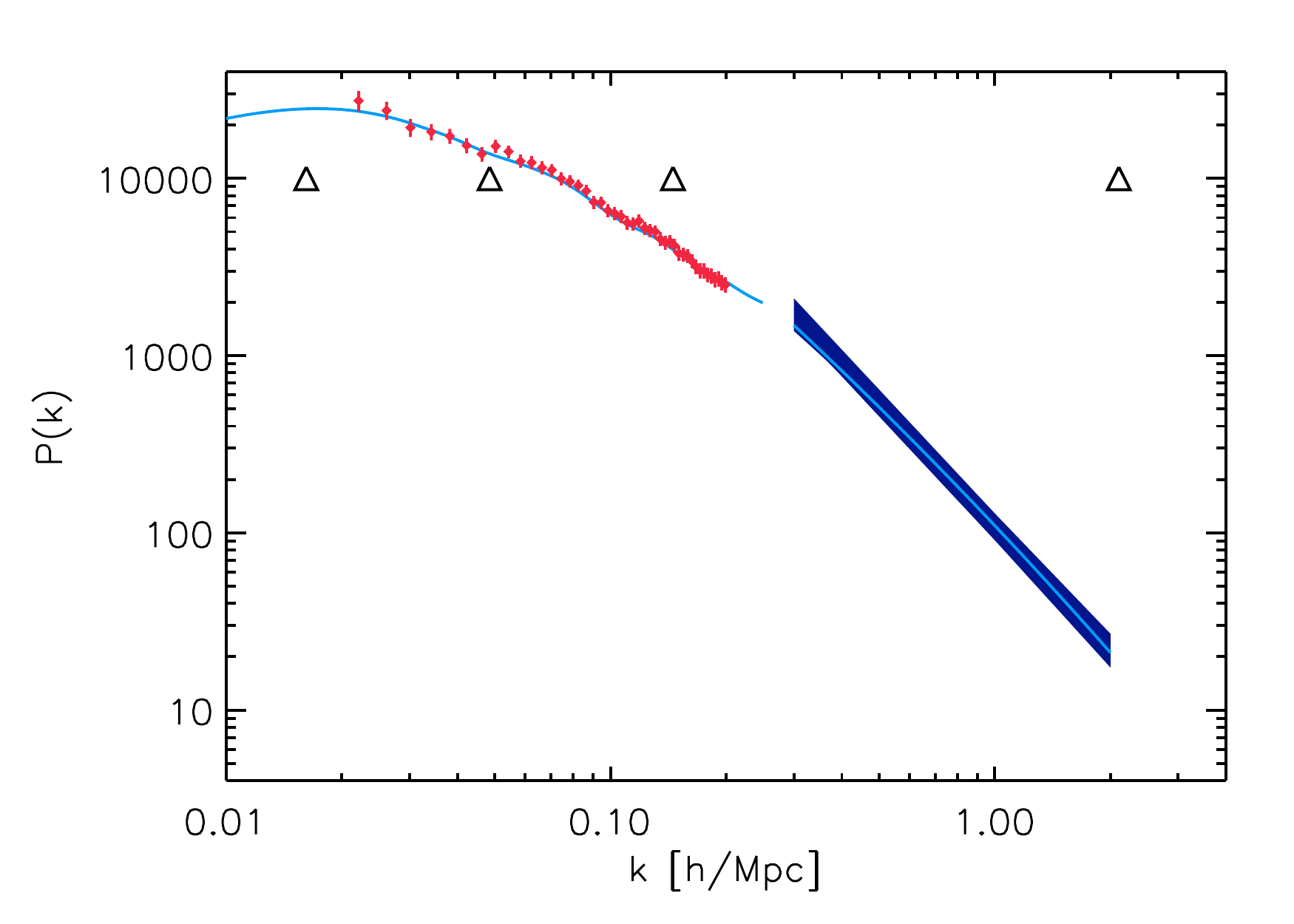}
\caption{Large-scale structure power spectrum in units of ($h$/Mpc)$^3$, showing knot placement  (triangles, arbitrary normalization) and cross-validation set-up. CV$_A$ is red and CV$_B$ is blue.  Red points represent the LRG power spectrum from Ref. \cite{Reidetal09}. The Lyman alpha  measurement is represented by a filled box encompassing the constraints on the observed flux power spectrum from Ref. \cite{McDonaldLya06}. The light blue line is a concordance LCDM model.} 
\label{fig:cvlss}
\end{figure}

We use 5  to 7 knots depending on the data set considered (see Table~\ref{tab:specs} for details; the locations of the knots in $k$ space are indicated in Figs.~\ref{fig:cvcmb} and \ref{fig:cvlss}). If the knot values were allowed infinite freedom and were set simply by minimizing the chi-square, in general the reconstruction would fit  features created by the random noise present in the data. It is therefore necessary to add a roughness penalty which we chose to be the integral of the second derivative of the spline function. The roughness penalty is weighted by a smoothing parameter:  
by increasing  the smoothing parameter the roughness penalty effectively reduces the degrees of freedom, disfavouring jagged functions that ``fit the noise''. 
\begin{table}
\caption{\label{tab:specs} The cross-validation set-up and the adopted number of knots for each data set used in the analysis.}
\begin{tabular}{cccc}
 DATA SET   & CV$_A$  & CV$_B$  & \# knots \\
 \hline
    WMAP5     &  yes\footnotemark[1] & yes \footnotemark[1]& $5$\\
    \hline
    QUaD      &    no&   yes & $6$\\
    ACBAR   &    yes    & no& $6$\\
    \hline
    SDSS DR7& yes & no & $6$\\
    Ly$\alpha$ & no&yes& $7$\\
    \hline
\end{tabular}
\footnotetext[1] {Following the choice as in Ref.~\cite{VerdePeiris08}, see Fig.~\ref{fig:cvcmb}}
\end{table}
In generic applications of smoothing splines, cross-validation is a rigorous statistical technique for choosing the optimal smoothing parameter. Cross-validation (CV) quantifies the notion that if the underlying  function has been correctly recovered, it should accurately predict new, independent data. To make the problem computationally manageable, we opt for a $n/2$--fold cross-validation, where $n$ is the number of data points. That is, the data set is split into two halves, say, $A$ and $B$. A Markov chain Monte Carlo (MCMC) parameter estimation analysis (for a given smoothing parameter) is carried out on one half of the data, finding the best fit model.  Then the log likelihood of the second half of the data given the best fit model for the first half,  CV$_{AB}$, is computed and stored. This is repeated by switching the roles of the two halves, obtaining CV$_{BA}$. The sum, CV$_{AB}$+CV$_{BA}$, gives the ``CV score" for that smoothing parameter. Finally, the smoothing parameter that best describes the entire data set is the one that minimizes the CV score. Table \ref{tab:specs} gives details of the implementation. Note that, as in Ref. \cite{VerdePeiris08}, the basic cosmological parameters ($\omega_b h^2$, $\Omega_ch^2$, $\Theta_A$, $\tau$) are varied in the MCMC as well as the values of the smoothing spline at the knots, which describe the primordial power spectrum. The MCMC is implemented with modified versions of the {\sf CAMB} \cite{camb} and {\sf COSMOMC} \cite{cosmomc} packages, with very stringent convergence criteria. Now we will describe our treatment of the data.

{\it CMB Data:} We use the $v3p2$ version of the WMAP5 likelihood function with standard options, with the temperature data divided into alternate (roughly equal signal-to-noise) $\ell$ bins for CV$_A$ and CV$_B$ respectively, exactly as in Ref. \cite{VerdePeiris08}.  The polarization data are always used in both CV cases. For  CV$_A$ we use the ACBAR bandpowers from Ref. \cite{ACBAR09} between $550\leq \ell \leq 1950$. For CV$_B$ we use the Pipeline 1 QUaD bandpowers between $569 \leq \ell \leq 2026$ from Ref. \cite{Quad09} (see Fig.~\ref{fig:cvcmb}).

{\it SDSS DR7 LRG Power Spectrum:} The LRG data are used in CV$_A$ with WMAP5 data. The data spans the range of wavenumbers $0.02 \leq k$ [$h$/Mpc] $ \leq 0.2$. The likelihood function we use is identical to that presented in Ref. \cite{Reidetal09} (see Fig.~\ref{fig:cvlss}).

{\it Lyman-alpha Constraints:} The Ly$\alpha$ data are used in CV$_B$ with WMAP5 data (see Fig.~\ref{fig:cvlss}). We use the publicly-released likelihood function by A. Slozar \cite{SlosarLya} to obtain Lyman-$\alpha$ forest constraints. For this likelihood to be valid, the model $P(k)$ must be well described by  a three parameter model of amplitude,  spectral slope and running 
at the Lyman-alpha forest scales {\it  i.e.} $0.3<k$ $[h/$Mpc$]<3$.  To check that this assumption holds in this $k$-range for our more general description of $P(k)$, we extrapolated the $P(k)$ from the Monte Carlo Markov chains of Ref. \cite{VerdePeiris08} to the Lyman-$\alpha$ scales and found that in this $k$-range the resulting spline can be well approximated by the prescription of Ref.~\cite{SlosarLya}. The residuals are at the percent level, well below the intrinsic  Lyman-$\alpha$ errors. With the more recent data sets we consider here, the approximation is expected to be even better.
\section{Results and Discussion}\label{sec:results}
\begin{figure*}[!htp]
\includegraphics[scale=0.48]{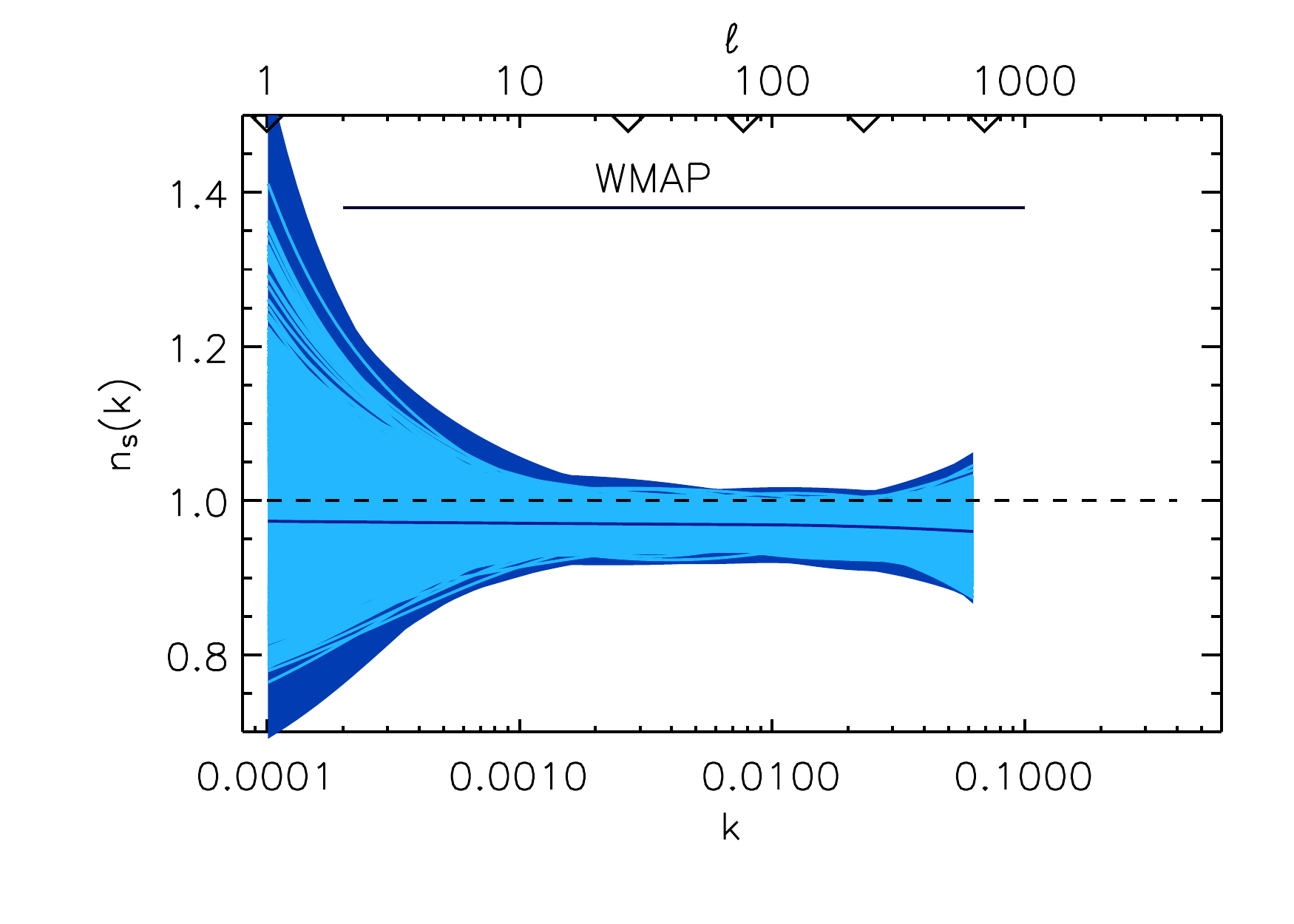} \hfill
\includegraphics[scale=0.48]{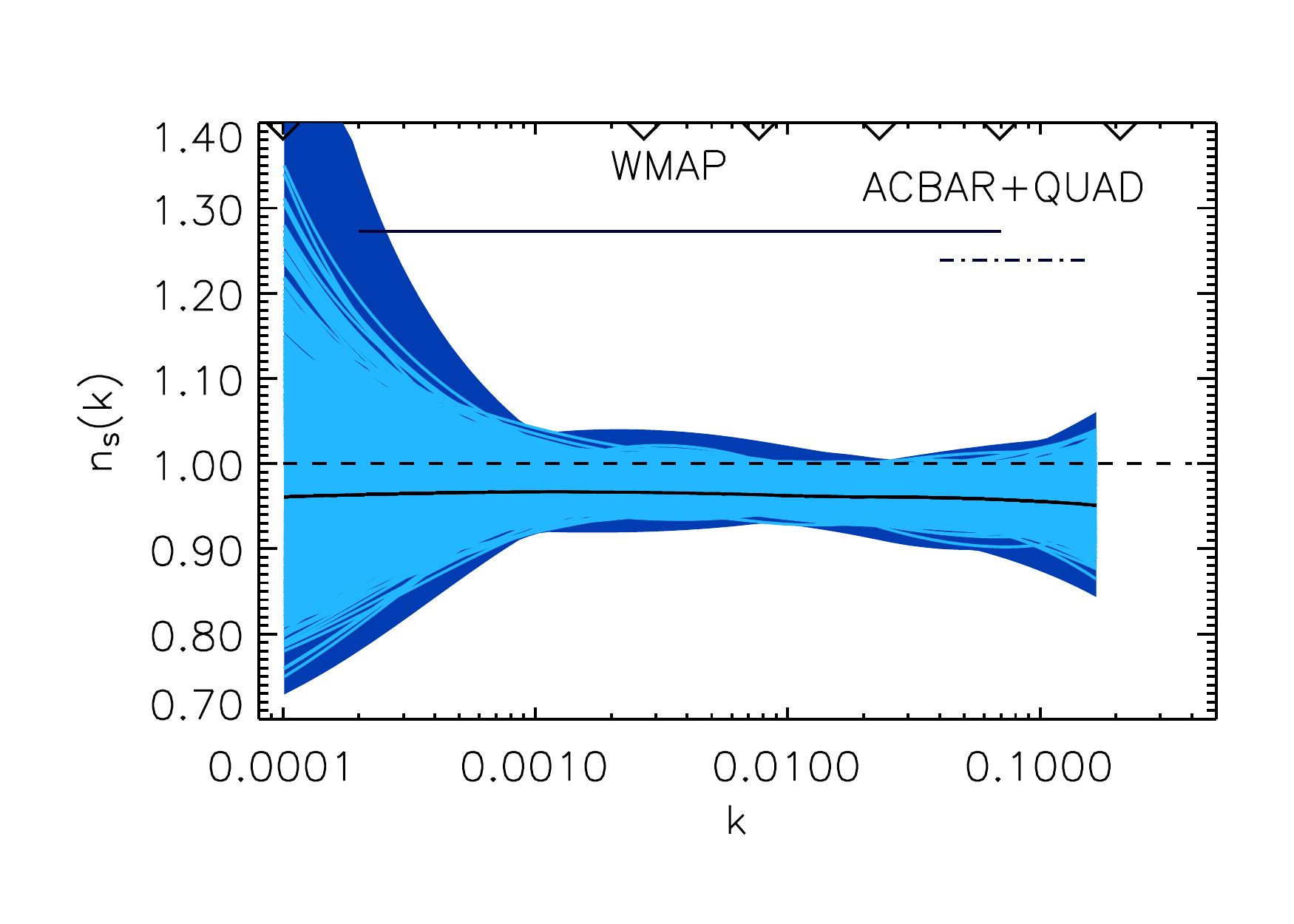} 
\includegraphics[scale=0.48]{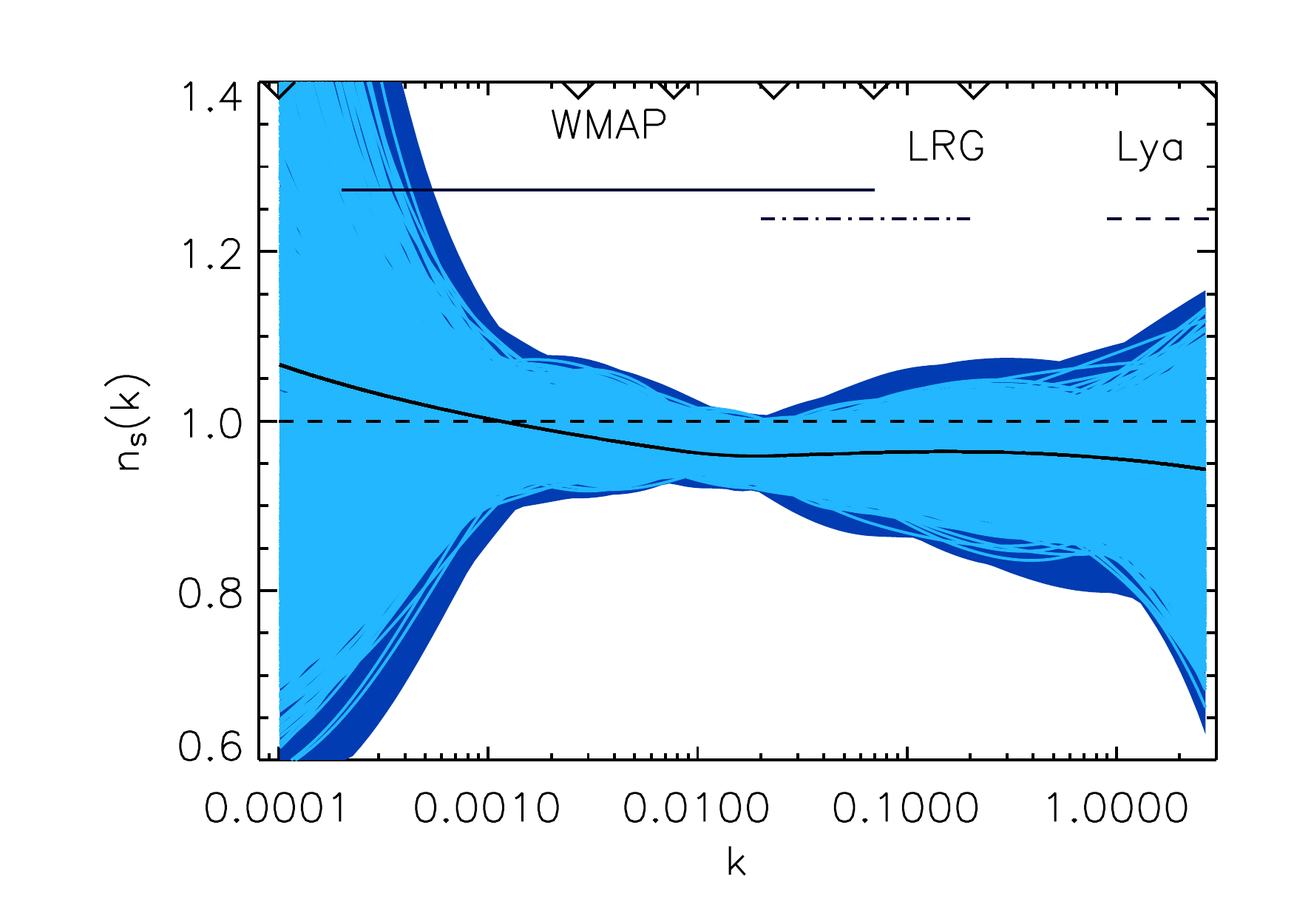} \hfill
\includegraphics[scale=0.48]{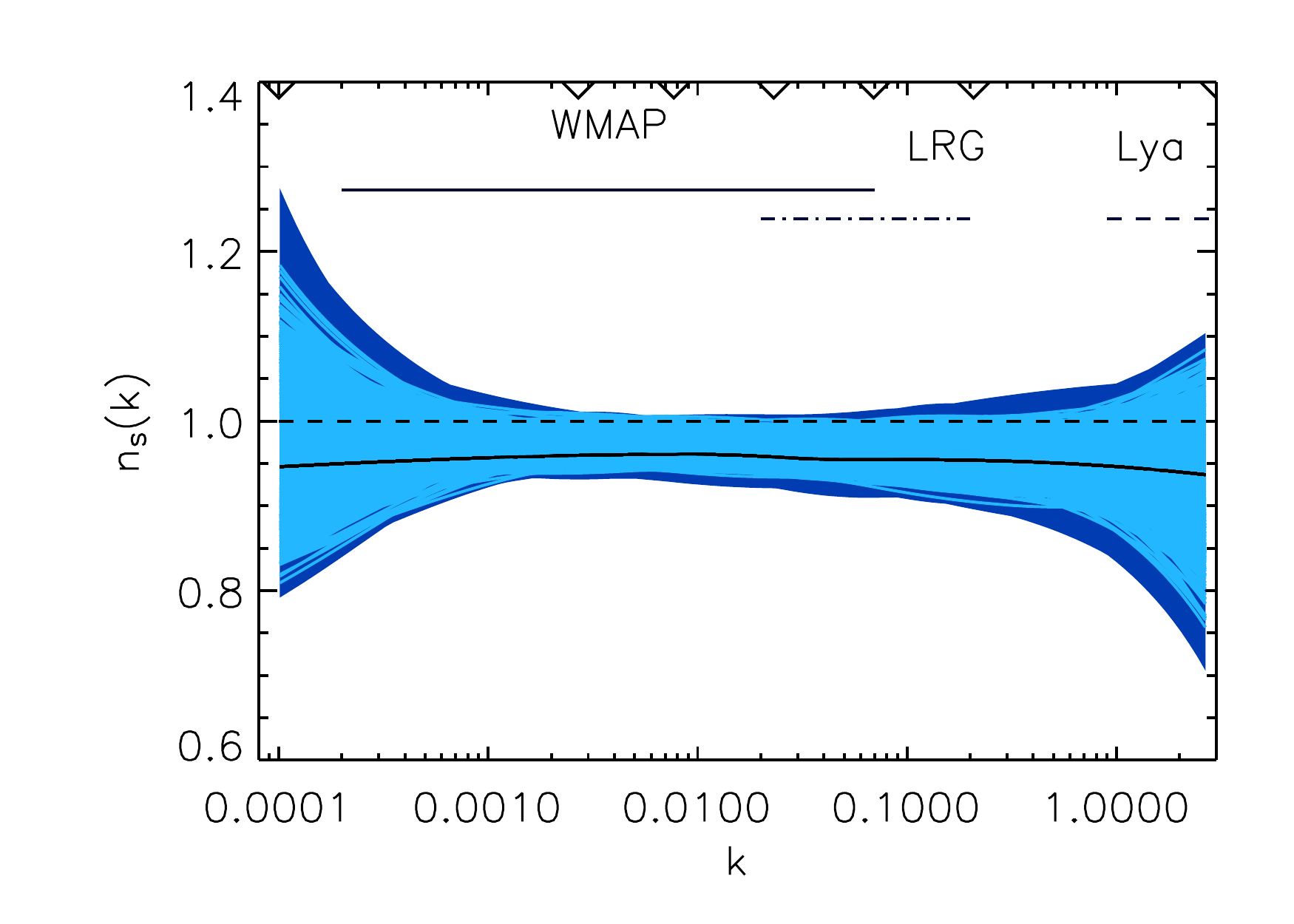} 
\caption{Reconstructed spectral index $n_s(k)$ for various data combinations:  WMAP5 with optimal penalty $\lambda^{\rm WMAP5}_{\rm opt}$ (top left), WMAP5$+$QUaD$+$ACBAR with optimal penalty $\lambda^{\rm CMB}_{\rm opt}$ (top right), and WMAP5$+$LRG$+$Ly$\alpha$ for two values of the  penalty (bottom, left and right). Dark  and light blue regions correspond to the best  95\% and 68\% reconstructions. The solid black line is the maximum likelihood fit. For comparison, the dashed line corresponds to a scale-invariant $P(k)$. See text for details.}
\label{fig:fourpanel}
\end{figure*}
Our main results are presented in Fig.~\ref{fig:fourpanel} for several data sets with increasing range in $k$: WMAP5 only, WMAP5 in combination with QUaD and ACBAR, and WMAP5 in combination with SDSS LRGs and Ly$\alpha$. We show the reconstructed $n_s (k)$ for ease of comparison with the standard power law results. However, the quantity that was actually reconstructed using cross-validation to find the optimal penalty is the power spectrum. 

The optimal penalty for WMAP5 $\lambda_{\rm opt}^{\rm WMAP5}$ is higher than what was found for WMAP3 by a factor of $25$, and is consistent with the optimal penalty for WMAP3 in combination with CMB data at smaller scales \cite{VerdePeiris08}.  The corresponding $n_s(k)$ is shown in the top left panel of Fig. \ref{fig:fourpanel}. 
The same optimal penalty is found for WMAP5 and WMAP5$+$QUaD$+$ACBAR ($\lambda^{\rm WMAP5}_{\rm opt}=\lambda^{\rm CMB}_{\rm opt}$), and the latter reconstruction is shown in the top right panel of Fig.~\ref{fig:fourpanel}.
For WMAP5$+$LRG$+$Ly$\alpha$, we find that CV becomes less sensitive to the value of the penalty, and the CV score dependence on the penalty flattens out at $\lambda_{\rm min}^{\rm WMAP5+LSS}=0.2\lambda^{\rm CMB}_{\rm opt}$.  While this may  indicate a preference for a less smooth $P(k)$, the data cannot distinguish between $\lambda_{\rm min}^{\rm WMAP5+LSS}$ and a penalty  an order of magnitude higher. The reconstructed $n_s(k)$  are shown in the left and right  bottom panels of Fig.~\ref{fig:fourpanel} for penalties $\lambda_{\rm min}^{\rm WMAP5+LSS}$ and $10 \lambda_{\rm min}^{\rm WMAP5+LSS}$, respectively. The dark and light blue regions enclose the best (ordered by likelihood) 95\% and 68\% reconstructions. The 95\% constraints are not significantly broader than the 68\% because the reconstructed spectra are simply more wiggly; they are not allowed by the data to deviate more from the best fit, consistently across scales.   

Cross-validation is a useful tool to check for indications of unidentified systematic biases in the data. For example, in Ref. \cite{VerdePeiris08} we found that the 3 year WMAP data (WMAP3) by itself favored a primordial power spectrum with a downward deviation from a power law at small scales (see Fig. 2 of Ref. \cite{VerdePeiris08}). However, this feature disappeared when combining WMAP3 with other data sets (see Figs. 3 and 5 of Ref. \cite{VerdePeiris08}) which overlapped WMAP3 on the scales corresponding to the feature -- an inconsistency suggestive of a small residual systematic effect in the high $\ell$ WMAP3 data.
Ref. \cite{Huffenbergeretal07} argued (based on considerations of frequency dependence) that the unresolved residual point source contribution to be subtracted from the raw $C_{\ell}$ should have been smaller  by  $28$\% -- and its  uncertainty increased by 60\% -- compared to the WMAP3 official values. To judge if smoothing spline cross-validation could give some insights on possible residual systematic errors, we investigated how the point source subtraction level should have been changed for the aforementioned downturn at small scales to disappear from the reconstructed power spectrum. We obtained a point source amplitude $\sim 20$\% lower than the  WMAP estimated value, which is tantalizingly close to the estimate of Ref. \cite{Huffenbergeretal07}.

In WMAP5, there is no longer any indication of deviations from a power-law primordial power spectrum, and the data require a smoother power spectrum (higher penalty) than WMAP3.

We find that WMAP5, CMB experiments at smaller scales, and the LRG power spectrum are all consistent with each other. With the addition of Ly$\alpha$ data, a lower penalty value is allowed. This could be a tentative indication of possible tension between Ly$ \alpha$ and the other data sets, but not a very significant one: there is a cancellation between the effect of penalty and the effect of the likelihood over a wide range of penalty values as shown in the bottom panels of Fig.~\ref{fig:fourpanel}). In addition, as LRG and Ly$\alpha$ scales do not overlap, we cannot exclude the possibility of a low-significance local feature in the power spectrum.

\begin{figure}[!htp]
\includegraphics[scale=0.48]{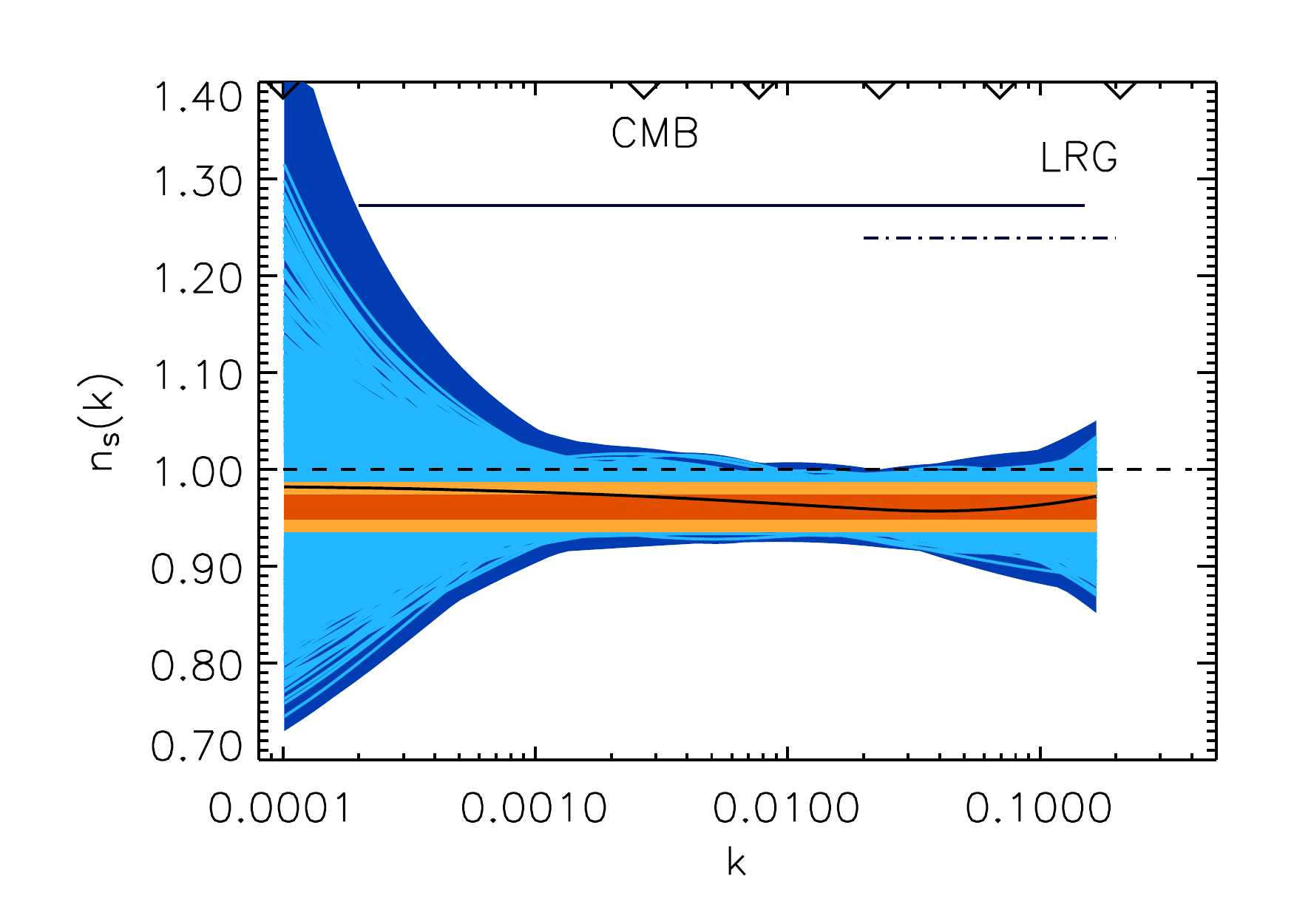} 
\caption{Reconstructed spectral index $n_s(k)$ from WMAP5, ACBAR, QUaD and SDSS  DR7 LRG  data with optimal penalty determined from cross-validation excluding Ly$\alpha$. The orange-red band shows  the 95\% and 68\% $n_s$ constraints \cite{Reidetal09} for WMAP5$+$LRG  data with a power-law prior.} 
\label{fig:cmbdr7}
\end{figure}

In Fig.~\ref{fig:cmbdr7} we show the reconstructed $n_s(k)$ for the CMB and LRG data, with optimal penalty $\lambda_{\rm opt}^{\rm WMAP5}$. The CV setup for WMAP5 is the same as before, LRGs are added in CV$_A$, and QUaD+ACBAR are included together in CV$_B$. We have excluded the Ly$\alpha$ data as it is the only non-overlapping data set. For comparison, we also show the 95\% and 68\% $n_s$ constraints \cite{Reidetal09} for WMAP5$+$LRG  data when a power-law spectral index is assumed to describe the shape of the primordial power spectrum. We see no evidence that any $k$-dependence of $n_s$ is necessary to describe the data in the CV reconstruction.  While $n_s=1$ is disfavoured, the significance of the departure from scale invariance is
weaker than when the ``inflation-motivated'' power law spectral index prior is adopted. 

This minimally-parametric reconstruction highlights how constraints relax when 
generic forms of $P(k)$ are allowed. While this reconstruction is in agreement with the inflationary prior, it illustrates that better data are needed to justify its adoption observationally. Forthcoming data from Planck will significantly reduce the current reliance on priors in our understanding of the shape of the primordial power spectrum. Future large-scale structure data and Planck will overlap over a decade in scale, offering extra consistency checks. Lyman alpha data, on the other hand, offer the potential to extend the lever arm by at least another decade. We hope that the results presented here will form a basis to  judge the robustness of our present knowledge when confronted with the precision measurements that are on the horizon.
\acknowledgments{HVP is supported by Marie Curie grant MIRG-CT-2007-203314 from the European 
Commission, a STFC Advanced Fellowship, and the Leverhulme Trust. LV acknowledges  MICINN grant AYA2008-03531 and FP7-IDEAS-Phys.LSS 240117 and thanks IoA Cambridge and DAMTP CTC for hospitality. We thank Matteo Viel for invaluable help with the interpretation of  Lyman alpha constraints and Beth Reid for crucial help with the SDSS LRG DR7 data and likelihood. Numerical computations were performed using the Darwin Supercomputer of the University of Cambridge High Performance Computing Service (http://www.hpc.cam.ac.uk/), provided by Dell Inc.~using Strategic Research Infrastructure Funding from the Higher Education Funding Council for England. We acknowledge the use of the Legacy Archive for Microwave Background Data (LAMBDA). Support for LAMBDA is provided by the NASA Office of Space Science. 
}
%


\end{document}